\documentclass[letterpaper,preprintnumbers,prd,twocolumn,nofootinbib,nobibnotes,showpacs]{revtex4}
\usepackage{amsfonts}
\usepackage{mathrsfs}
\usepackage{epsfig}
\usepackage{graphicx}%
\usepackage{dcolumn}
\usepackage{amsmath}

\makeatletter
\def\btt#1{\texttt{\@backslashchar#1}}%
\DeclareRobustCommand\bblash{\btt{\@backslashchar}}%
\makeatother
\begin{document}

\title{Modified Entropic Force}
\author{Changjun Gao}\email{gaocj@bao.ac.cn}\affiliation{The
National Astronomical Observatories, Chinese Academy of Sciences}
\affiliation{{Key Laboratory of Optical Astronomy, NAOC, CAS,
Beijing, 100012
}}
\affiliation{{Kavli Institute for Theoretical
Physics China, CAS, Beijing 100190, China }}
\date{\today}

\begin{abstract}
The theory of statistical thermodynamics tells us the equipartition law of energy does not hold in the limit of very low temperatures. It is found the Debye model is very successful in explaining the experimental results for most of the solid objects. Motivated by this fact, we modify the entropic force formula which is proposed very recently.
Since the Unruh temperature is proportional to the strength of gravitational field, so the modified entropic force formula is an extension of the Newtonian gravity to weak field. On the contrary, General Relativity extends Newtonian gravity to strong field case. Corresponding to Debye temperature, there exists a Debye acceleration $g_D$. It is found the Debye acceleration is $g_D=10^{-15}\textrm{N}\cdot \textrm{{kg}}^{-1}$. This acceleration is very much smaller than the gravitational acceleration $10^{-4}\textrm{N}\cdot \textrm{{kg}}^{-1}$ which is felt by the Neptune and the gravitational acceleration $10^{-10}\textrm{N}\cdot \textrm{{kg}}^{-1}$ felt by the Sun. Therefore, the modified entropic force can be very well approximated by the Newtonian gravity in the solar system and in the Galaxy. With this Debye acceleration, we find the current cosmic speeding up can be explained without invoking any kind of dark energy.
\end{abstract}


\maketitle

\section{Introduction}
Recently, Verlinde \cite{verlinde} makes one interesting proposal that gravity may be explained as an entropic force
caused by the changes in the information associated with the positions of material bodies.
The proposal is very interesting in the following two aspects. In the first place, with the assumption of the entropic force together with the Unruh temperature
\cite{unruh:1976}, Verlinde is able to derive the second law of Newtonian mechanics. Secondly, the
assumption of the entropic force together with the holographic principle \cite{holography} and the equipartition
law of energy can lead to the Newtonian law of gravitation. Similar discoveries are
also made by Padmanabhan \cite{pad:2009}. Padmanabhan finds that using the equipartition law of energy for
the horizon and the thermodynamic relation $S=E/2T$,
one can obtain the Newtonian law of gravity. Here $S$ and $T$ are thermodynamic entropy and
the temperature associated with the horizon. $E$ is the active gravitational mass which produces
the gravitational field in the spacetime \cite{pad:2004}. Actually, Jacobson has derived the Einstein's equations by employing
the Clausius relation and the equivalence principle in 1995 \cite{Jac:1995}.

Using the entropic force, Shu and Gong \cite{shu}, Cai, Cao and Ohta
\cite{cai} derived the Friedmann equation nearly simultaneously
while with different methods. On the other hand, using this entropic
force, Smolin \cite{smolin} derived the Newtonian gravity in loop
quantum gravity. M$\ddot{\textrm{{a}}}$kel$\ddot{\textrm{{a}}}$
\cite{makela} pointed out that Verlinde's entropic force is actually
the consequence of a specific microscopic model of spacetime
\cite{mak3}. Similar ideas were applied to the construction of
holographic actions from black hole entropy by Caravelli and Modesto
\cite{car} . Li and Wang \cite{miaoli} showed that the holographic
dark energy \cite{li:2004} can be derived from the entropic force
formula. \footnote{Same observation was disclosed by Johannes
Koelman in his blog. (see http://www.scientificblogging.com-hammock
physicist-it bit how get rid dark energy)}

We note that statistical thermodynamics reveals the equipartition law of energy does not hold in the very low temperatures. It is found the Debye model is very successful in explaining the experimental results for most of the solid objects. In particular, many experiments on Debye model indicate that the lower the temperature, the better the consistency is.

On the other hand, the quantum theory of black holes finds black holes satisfy the first law of
thermodynamics \cite{fourlaws}. Furthermore, the Unruh effect indicates that a gravitational field corresponds a thermal bath with temperature of $T$. So every gravitational system may be physically corresponding to a statistical thermodynamics system. Since the equipartition law of energy does not hold in low temperatures for thermodynamic system, we anticipate the entropic force formula should be modified for the very weak gravitational fields.

Actually, the strength of gravitational fields are $10^{1}\ \textrm{N}\cdot \textrm{{kg}}^{-1}$ on the earth, $10^{-4}\textrm{N}\cdot \textrm{{kg}}^{-1}$ for the solar system (felt by the Neptune) and $10^{-10}\textrm{N}\cdot \textrm{{kg}}^{-1}$ for the Galaxy (felt by the Sun). In one word, the strength of gravitational field becomes smaller and smaller with the increasing of scales. Therefore, it may be plausible that the entropic force (or Newtonian gravity) takes a different form in the background of an extreme weak gravitational field.

In this paper, motivated by the Debye's model in thermodynamics, we extend the entropic force formula. With this modification, we find
the resulting entropic force can speed up the expansion of the Universe.
We shall use the system of units with $G=c=\hbar=k=1$ and the metric signature $(-,\ +,\ +,\ +)$ throughout the paper.
\section{Modified entropic force}
Following \cite{verlinde}, we consider a compact spatial
region $V$ with a compact boundary $B$ which is a sphere.
The compact boundary $B$ plays the role of a holographic screen. The number $N$ of bits for freedom on the
screen is taken to be equal to the area $A$ of the screen:
\begin{eqnarray}
 \label{eq:na}
 N=A\;.
 \end{eqnarray}
Let $g$ denotes the proper gravitational acceleration on the screen which is produced
by the matter sources inside the screen $B$. Then the Unruh effect tells us the gravitational field would correspond to a system filled with thermal
gas with the temperature of $T$ for a proper observer:
\begin{eqnarray}
 \label{eq:unruh}
 T=\frac{g}{2\pi}\;.
 \end{eqnarray}
So according to the equipartition law of energy, the total energy inside the screen is
\begin{eqnarray}
 \label{eq:energy}
 E=\frac{1}{2}NT\;.
 \end{eqnarray}
Verlinde assumes the energy is equal to the gravitational mass $M$ inside the screen:
\begin{eqnarray}
 \label{eq:mass}
 E=M\;.
 \end{eqnarray}
Combining above equations, we can obtain the spherically symmetric and static gravitational field equations:
\begin{eqnarray}
 \label{eq:mass}
 M=\frac{1}{2}\cdot4\pi r^2\cdot\frac{g}{2\pi}\;,
 \end{eqnarray}
 namely,
 \begin{eqnarray}
 \label{eq:mass}
 g=\frac{M}{r^2}\;.
 \end{eqnarray}
 It is exactly the Newtonian law of gravity. Statistical thermodynamics tells us the equipartition law of energy fails in
 the very low temperatures. It is found the Debye model is very successful for most of solid objects.
 Since a gravitational system may be physically corresponding to a statistical thermodynamics system and the strength of gravitational field
 becomes smaller and smaller with the increasing of cosmic scales, we anticipate the Newtonian gravity may be modified on the very large scale of the Universe.

 Motivated by this point,
 we modify the equipartition law of energy as follows:
 \begin{eqnarray}
 \label{eq:debeye}
 E=\frac{1}{2}NT\mathfrak{D}\left(x\right)\;,
 \end{eqnarray}
 where the Debye function is defined by
 \begin{eqnarray}
 \label{eq:dfunction}
 \mathfrak{D}\left(x\right)\equiv\frac{3}{x^3}\int_0^{x}\frac{y^3}{e^{y}-1}dy\;.
 \end{eqnarray}
$x$ is related to the temperature $T$ as follows:
\begin{eqnarray}
 \label{eq:tem}
x\equiv\frac{T_{D}}{T}\;.
 \end{eqnarray}
$T_{D}$ is the Debye critical temperature. For a gravitational system, $T_{D}$ is related to the critical
gravitational acceleration. So we obtain the modified entropic force from Eq.~(\ref{eq:debeye})
as follows

\begin{eqnarray}
 \label{eq:gg}
 g=\frac{M}{r^2}\cdot\frac{1}{\mathfrak{D}\left(x\right)}\;,
 \end{eqnarray}
with
\begin{eqnarray}
 x\equiv\frac{g_D}{g}\;.
 \end{eqnarray}
The Debye acceleration is related to the Debye temperature
\begin{eqnarray}
 g_D={2\pi T_D}\;.
 \end{eqnarray}
We note that when $x\ll1$, i.e. in the limit of strong gravitational field,
we have
\begin{eqnarray}
  \mathfrak{D}\left(x\right)=1\;.
 \end{eqnarray}
So the Newtonian gravity is recovered. On the other hand, in the limit of weak gravitational field, i.e, $x\gg1$,
we have
\begin{eqnarray}
  \mathfrak{D}\left(x\right)=\frac{\pi^4}{5x^3}\;,
 \end{eqnarray}
 and
 \begin{eqnarray}
 \label{eq:mod}
 g=\left(\frac{5Mg_D^3}{\pi^4}\right)^{\frac{1}{4}}\cdot\frac{1}{\sqrt{r}}\;.
 \end{eqnarray}
 In this limit, the gravitational field is very different from Newtonian gravity. In the next, let's
 look for the value of $g_D$ from the considerations of cosmological physics.
Using the method of Shu and Gong \cite{shu}, we obtain the following equation
\begin{eqnarray}
 \label{eq:eq}
 4\pi\left(\rho+p\right)=-\left(\dot{H}-\frac{K}{a^2}\right)\left[-2 \mathfrak{D}\left(x\right)+\frac{3x}{e^{x}-1}\right]\;.
 \end{eqnarray}
 Here $\rho, \ p$ are the total energy density and total pressure of cosmic fluids, respectively. $a,\ H$ are the scale factor and Hubble parameter, respectively. $K=0,\ +1,\ -1$ describe the three kind of geometry of the Universe. Dot denotes the derivative with respect to the comic time $t$.
 Taking into account the Hawking temperature $T$ for the Universe \cite{cait1,cait2}
 \begin{eqnarray}
  T=\frac{H}{2\pi}\;,
 \end{eqnarray}
 we find $x$ can be rewritten as
 \begin{eqnarray}
 x\equiv\frac{H_D}{H}\;.
 \end{eqnarray}
Thus the critical Hubble parameter $H_D$ is exactly the Debye acceleration $g_D$:
\begin{eqnarray}
 H_D={g_D}\;.
 \end{eqnarray}
Eq.~(\ref{eq:eq}) together with the energy conservation equation
\begin{eqnarray}
\dot{\rho}+3H\left(\rho+p\right)=0\;,
\end{eqnarray}
govern the evolution of the Universe. Using these two equations, we find the current cosmic acceleration can be interpreted while
without invoking any kind of dark energy. To show this point, we may investigate the cosmic evolution sourced by dark matter and baryon matter.
Observations reveal that the Universe is highly flat in space. So in the following we shall put $K=0$.
Define
\begin{eqnarray}
u\equiv\ln a\;.
\end{eqnarray}
Combining Eq.~(16) and Eq.~(20) we obtain the Friedmann equation:
\begin{eqnarray}
8\pi\rho=\int_{}^{}{}3\left[-2\mathfrak{D}\left(x\right)+\frac{3x}{e^{x}-1}\right]dH^2\;.
\end{eqnarray}

Let
\begin{eqnarray}
h\equiv\frac{H}{H_0}\;,\ \ \Omega_{m0}\equiv\frac{\rho_{m0}}{\rho_0}\;,\ \  H_D=\zeta H_0\;,
\end{eqnarray}
then we have
\begin{eqnarray}
{\Omega_{m0}}e^{-3u}=\int_{\infty}^{x}\left[4\mathfrak{D}\left(x\right)-\frac{6x}{e^{x}-1}\right]\cdot\frac{\zeta^2}{x^3}dx\;,
\end{eqnarray}
and
\begin{eqnarray}
x=\frac{\zeta}{h}\;.
\end{eqnarray}
Here $H_0, \ \rho_0,\ \rho_{m0}$ are the Hubble parameter, total energy density and matter energy density ratio
for the present-day Universe. $\zeta$ is a dimensionless free parameter. The
standard cosmological model, e.g. as in Komatsu et al.
\cite{Komatsu5yrWMAP}, predicts a present matter density ratio $\Omega_{m0}=0.25$. Using this result and taking $\zeta=10^{-5}$, we plot the dimensionless Hubble parameter $h$ via redshifts $z$ in Fig.~1 for our model and the standard $\Lambda \textrm{CDM}$ model. We find they have nearly
the same behavior.

\begin{figure}
\includegraphics[width=6.5cm]{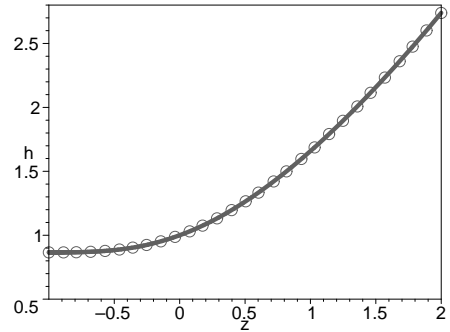}
\caption{The dimensionless Hubble parameter $h$ with redshift $z$. The solid line is for the standard $\Lambda \textrm{CDM}$ model. The circled line is for our model.} \label{fig:hz}
\end{figure}
In order to show our model can lead to the cosmic acceleration, we plot the evolution of deceleration parameter $q$:
 \begin{eqnarray}
q\equiv-\left(1+\frac{d\ln h}{du}\right)\;,
\end{eqnarray}
in Fig.~2. We find the model predicts nearly the same transition redshift $z_T=0.8$ as the standard $\Lambda \textrm{CDM}$ model.
\begin{figure}
\includegraphics[width=6.5cm]{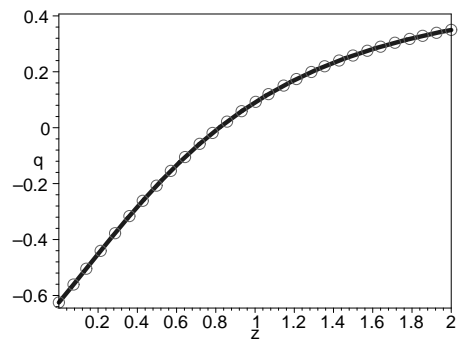}
\caption{The deceleration parameter $q$ with redshift $z$. The solid line is for the standard $\Lambda \textrm{CDM}$ model. The circled line is for our model.} \label{fig:qz}
\end{figure}

At higher redshifts (large $H$), we have $x\ll 1$. Then Eq.~(\ref{eq:eq}) restores to the standard equation in General Relativity. So the physics at and before the radiation dominated epoch is not modified. Finally, let's calculate the Debye acceleration $g_D$:
 \begin{eqnarray}
g_D=\zeta H_0=10^{-15}\textrm{N}\cdot \textrm{{kg}}^{-1}\;.
\end{eqnarray}
It is the order of $10^{-11}$ of the gravitational field strength in the solar system and the order of $10^{-5}$ for the Galaxy system. So Eq.~(\ref{eq:gg}) can be perfectly
written as the Newtonian law of gravity in the solar system and the Galaxy scales. To show this point in great detail, we may observe the modified Poisson's equation which is given by:
 \begin{eqnarray}
 \label{eq:poi}
 \nabla\cdot\left[\mathfrak{D}\left(x\right)\nabla\Phi\right]=4\pi\rho\;,
 \end{eqnarray}
 with
 \begin{eqnarray}
  x=\frac{g_D}{|\nabla\Phi|}\;.
 \end{eqnarray}
Here $\rho$ is the energy density of matter sources. $\Phi$ is the Newtonian gravitational potential. When $x\ll 1$, i.e. in the limit of strong gravitational field, we have $\mathfrak{D}=1$ which is for the Newtonian gravity. So the deviation of Eq.~(\ref{eq:poi}) from Newtonian gravity is embodied by the Debye function. In Fig.~3, we plot the values of Debye function in Galaxy. The strength of gravitational field in the Galaxy is around the order of $10^{5}$ of the Debye acceleration $g_D$. It then follows that the Debye function $\mathfrak{D}\simeq 1$ in the Galaxy such that the Newtonian gravity emerges as a very good approximation.

\begin{figure}
\includegraphics[width=6.5cm]{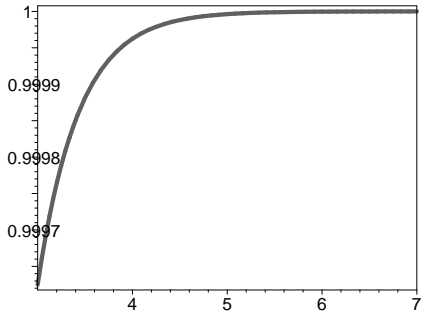}
\caption{The values of Debye function $\mathfrak{D}$ with the strength of gravitational field $\log\left({|\nabla\phi|}/{g_D}\right)$ in Galaxy. } \label{fig:qz}
\end{figure}

\section{conclusion and discussion}
Statistical thermodynamics tells us the equipartition law of energy does not hold in
the limit of temperature approaching zero. It is found that the Debye model is very successful in interpreting the physics of very low temperature system for most of the solid objects. Since a gravitational system can also be taken as a thermodynamical system, we expect the equipartition law of energy for gravitational system shall break down in the limit of very low temperature (or very low gravitational field strength). Motivated by Debye model, we modified the entropic force. We find the modification can interpret the current acceleration of the Universe while without invoking any kind of dark energy.

The model is closely related the Debye acceleration $g_D$. We find $g_D=10^{-15}\textrm{N}\cdot \textrm{{kg}}^{-1}$ is consistent with cosmic observations very well. This gravitational field strength is very much smaller than the gravitational fields, by orders of $10^{11}$ in solar system and by orders of $10^{5}$ in the Galaxy. Therefore, the modified entropic force is not inconsistent with the solar system tests and the Galaxy scale observations. Then how to derive the Debye model from the specific microscopic model of spacetime? We leave this problem to future study.

\acknowledgments
I am grateful to the referees for the insightful
comments and suggestions, which have allowed me to improve this paper significantly. I sincerely thank Dr. Savvas Nesseris for stimulating and
illuminating discussions which help me find two typographical errors in Eq.~(22) and Eq.~(24). Special thanks go to Prof. Miao Li, Dr. Fuwen. Shu, Dr. Xin. Zhang and Dr. Yu Tian for many helpful discussions. This work is
supported by the National Science Foundation of China under the
Key Project Grant 10533010, Grant 10575004, Grant
10973014, and the 973 Project (No. 2010CB833004).

\newcommand\AL[3]{~Astron. Lett.{\bf ~#1}, #2~ (#3)}
\newcommand\AP[3]{~Astropart. Phys.{\bf ~#1}, #2~ (#3)}
\newcommand\AJ[3]{~Astron. J.{\bf ~#1}, #2~(#3)}
\newcommand\APJ[3]{~Astrophys. J.{\bf ~#1}, #2~ (#3)}
\newcommand\APJL[3]{~Astrophys. J. Lett. {\bf ~#1}, L#2~(#3)}
\newcommand\APJS[3]{~Astrophys. J. Suppl. Ser.{\bf ~#1}, #2~(#3)}
\newcommand\JHEP[3]{~JHEP.{\bf ~#1}, #2~(#3)}
\newcommand\JCAP[3]{~JCAP. {\bf ~#1}, #2~ (#3)}
\newcommand\LRR[3]{~Living Rev. Relativity. {\bf ~#1}, #2~ (#3)}
\newcommand\MNRAS[3]{~Mon. Not. R. Astron. Soc.{\bf ~#1}, #2~(#3)}
\newcommand\MNRASL[3]{~Mon. Not. R. Astron. Soc.{\bf ~#1}, L#2~(#3)}
\newcommand\NPB[3]{~Nucl. Phys. B{\bf ~#1}, #2~(#3)}
\newcommand\PLB[3]{~Phys. Lett. B{\bf ~#1}, #2~(#3)}
\newcommand\PRL[3]{~Phys. Rev. Lett.{\bf ~#1}, #2~(#3)}
\newcommand\PR[3]{~Phys. Rep.{\bf ~#1}, #2~(#3)}
\newcommand\PRD[3]{~Phys. Rev. D{\bf ~#1}, #2~(#3)}
\newcommand\RMP[3]{~Rev. Mod. Phys.{\bf ~#1}, #2~(#3)}
\newcommand\SJNP[3]{~Sov. J. Nucl. Phys.{\bf ~#1}, #2~(#3)}
\newcommand\ZPC[3]{~Z. Phys. C{\bf ~#1}, #2~(#3)}
\newcommand\CQG[3]{~Class. Quant. Grav.{\bf ~#1}, #2~(#3)}
\newcommand\CMP[3]{~Commun. Math. Phys.{\bf ~#1}, #2~(#3)}

\end{document}